\documentclass[submitting]{nst}

\usepackage{subfigure,dcolumn}
\usepackage{epstopdf}
\usepackage{mhchem}
\usepackage{upgreek}

\usepackage{graphicx}
\usepackage{bm}
\usepackage{color}
\usepackage{booktabs,longtable}
\usepackage{multirow}
\usepackage{CJK}
\usepackage{ltxtable}
\usepackage{rotating}

\begin{document}

\title{Searching for single-particle resonances with the Green's function method}
\thanks{Supported by the National Natural Science Foundation of China (No.~U2032141), and the Natural Science Foundation of Henan Province (No.~202300410479 and No.~202300410480), the Foundation of Fundamental Research for Young Teachers of Zhengzhou University (No.~JC202041041), and the Physics Research and Development Program of Zhengzhou University (No.~32410217).}

\author{Ya-Tian Wang}
\affiliation{School of Physics and Microelectronics, Zhengzhou University, Zhengzhou 450001, China}

\author{Ting-Ting Sun}
\email[Corresponding author: ]{ttsunphy@zzu.edu.cn}
\affiliation{School of Physics and Microelectronics, Zhengzhou University, Zhengzhou 450001, China}

\begin{abstract}

Single-particle resonances in the continuum are crucial for studies of exotic
nuclei. In this study, the Green's function approach is employed to search for
single-particle resonances based on the relativistic-mean-field~model. Taking
$^{120}$Sn as an example, we identify single-particle resonances and determine the
energies and widths directly by probing the extrema of the Green's functions. In
contrast to the results found by exploring for the extremum of the density of states proposed in
our recent study~[Chin. Phys. C, 44:084105 (2020)], which has proven to be very
successful, the same resonances as well as very close energies and widths are
obtained. By comparing the Green's functions plotted in different coordinate space
sizes, we also found that the results very slightly depend on the space size.
These findings demonstrate that the approach by exploring for the extremum of the Green's function
is also very reliable and effective for identifying resonant states, regardless of whether they are wide or
narrow.

\end{abstract}

\keywords{Single-particle resonances, extrema of Green's functions, relativistic-mean-field theory}

\maketitle

\section{Introduction}
\label{sec:intr} Recently, explorations for single-particle resonances are
attracting increasing attention because of their significant role in studies of
exotic nuclei. Many exotic phenomena such as halos~\cite{PRL1985Tanihata_55_2676},
deformed halos~\cite{PRL2014Kobayashi}, and giant
halos~\cite{PRL1998MengJ_80_460,PRC2002MengJ_65_041302,PRC2006Grasso_74_064317} are
explained by the occupations of valence neutrons in the continuum. For example, giant
halos predicted in neutron-rich Zr and Ca isotopes are caused by valence neutrons
scattered to the continuum and occupying $p$
orbitals~\cite{PRL1998MengJ_80_460,PRC2002MengJ_65_041302}, and the possible deformed halos
in $^{40,42}$Mg and $^{22}$C are mainly caused by the occupations of single-particle
states around the Fermi
surface~\cite{PRC2010Zhou_82_011301,PRC2012Li_85_024312,PLB2018SunXX_785_530}. Numerous studies have shown that, in weakly bound exotic nuclei with very small gaps,
between the Fermi surface and the continuum threshold, the valence nucleons can be
scattered to the continuum effortlessly by pairing correlations. Halos can be formed
if the valence nucleons occupy an orbit with a small angular momentum $l$, which can
contribute a large radius~\cite{PRC1996Dobaczewski_53_2809,NPA1998JMeng_635_3}.

To explore single-particle resonances, researchers have developed a series of
approaches. One technique starts from scattering theory, such as K-matrix
theory~\cite{PRC1991Humblet_44_2530}, S-matrix
theory~\cite{Book1972Taylor-ScatteringTheor,PRC2002CaoLG_66_024311}, R-matrix
theory~\cite{PRL1987Hale_59_763,PR1947Wigner_72_29}, the Jost function
approach~\cite{PRL2012LuBN_109_072501,PRC2013LuBN_88_024323}, and the scattering phase
shift method~\cite{PRC2010LiZP_81_034311,SCP2010Li_53_773}. Meanwhile, approaches
for bound states are also widely used; these include the real stabilization
method \cite{PRA1970Hazi_1_1109,NPA2004Hagino_735_55}, the complex scaling
method \cite{PRC1986Gyarmati_34_95,PRL1997Kruppa_79_2217,PRC2010JYGuo_82_034318,NST2016ZMNiu},
the analytical continuation of the coupling constant
method \cite{PRC2004ZhangSS_70_034308,PLB2014SSZhang_730_30}, the complex
momentum representation method~\cite{PRC2006Hagen_73_034321,PRL2016Li_117_062502},
and the complex-scaled Green's
function method~\cite{EPJA2017Shi_53_40}.

The Green's
function approach~\cite{PRB1992Tamura_45_3271,PRA2004Foulis_70_022706}, which has wide applications in various fields of
physics~\cite{CPL2019AHorri,CPL2019SHJia},~has
also been demonstrated to be very effective for studying the continuum and
single-particle resonant states. Owing to its advantages, such as its being able to handle
bound states and the continuum uniformly by using the density of states (DOS) tool, the resonance
energies and widths can be determined easily, the asymptotic behavior of the spatial
extended density in weakly bound nuclei can be properly described, and, most
importantly, it can be combined with different nuclear models very conveniently. The Green's
function method has yielded significant achievements in nuclear physics in investigating the
effects of the continuum on the properties of atomic nuclei. For example, to
describe the continuum exactly in exotic nuclei near the drip line and study the
possible effects of the continuum on the properties of the ground state, Zhang \emph{et al.}
developed the self-consistent continuum Skyrme Hartree--Fock--Bogoliubov~(HFB)
theory~\cite{PRC2011ZhangY_83_054301,PRC2012YZhang_86_054318}, based on which a series
of research projects have been conducted. These include establishment of the energies and wave functions for
single-particle canonical bound and resonant states with different space
sizes~\cite{PRC2019QuXY_99_014314}, study of the possible impacts of mean-field and pairing on
the resonances~\cite{SC2019QuXY, PRC2020ZhangY}, and the extension to the odd-$A$ systems
by including the blocking effect~\cite{PRC2019SunTT_99_054316}. To explore the contribution
of the continuous spectrum to nuclear collective excitations, Matsuo applied the HFB
Green's function~\cite{SJNP1987Belyaev_45_783} to the quasiparticle random-phase
approximation \cite{NPA2001Matsuo_696_371,PTPS2002Matsuo_146_110}, enabling further study of the
collective excitations coupled to the
continuum~\cite{PRC2005Matsuo_71_064326,PRC2009Mizuyama_79_024313,PRC2010Matsuo_82_024318,PRC2011Shimoyama_84_044317},
microscopic structures of monopole pair vibrational modes and associated two-neutron
transfer amplitudes~\cite{PRC2013Shimoyama_88_054308}, and neutron capture
reactions.

Given the great successes that the Green's function method achieved in the nonrelativistic
framework, it is naturally applied in covariant density functional
theory~\cite{PPNP2006MengJ_57_470,PPNP1996PRing_37_193,PR2005Vretenar_409_101},
which has been demonstrated to be a powerful tool in researching various nuclear systems and
properties, such as superheavy
nuclei~\cite{PPNP2007Sobiczewski,PRC2012WangN_85_041601,PRC2017ZhangW_96_054308},
pseudospin
symmetry~\cite{PhysRep2015HZLiang_570_1,JPG2017Lu_44_125104,PRC2017Sun_96_044312},
hypernuclei~\cite{PRC2011BNLu_84_014328,PRC2016TTSun_94_064319,PRC2018LiuZX_98_024316},
and neutron stars~\cite{NST2017BJCai,CPC2018SunTT_42_025101}. In
Refs.~\cite{PRC2009Daoutidis_80_024309,PRC2010DYang_82_054305}, the relativistic
continuum random-phase
approximation theory is developed by adopting the Green's function of the Dirac
equation~\cite{PRB1992Tamura_45_3271} to investigate collective excitations. In
Ref.~\cite{PRC2014TTSun_90_054321}, we introduced the Green's function approach to
the relativistic-mean-field (RMF) model and studied single-particle resonances for the first time. Later, this approach
was further extended to studies of single-particle resonances of
protons~\cite{JPG2016TTSun_43_045107}, hyperons~\cite{PRC2017Ren_95_054318}, and those in
deformed nuclei with a quadrupole-deformed Woods--Saxon
potential~\cite{PRC2020Sun_101_014321}. In addition,
the pseudospin symmetries hidden in resonant states were also
investigated by applying the Green's function method
\cite{PRC2019Sun_99_034310}. In Ref.~\cite{Sci2016Sun_46_12006}, to study
the halo structures in neutron-rich nuclei, we further included the pairing correlation
and introduced the Green's function approach to the continuum relativistic
Hartree--Bogoliubov theory.

In our previous studies of single-particle resonant
states~\cite{PRC2014TTSun_90_054321,JPG2016TTSun_43_045107,PRC2017Ren_95_054318}, the
resonant states are determined by comparing the DOS of particles in the
mean field to those for free particles. In this framework, resonance energies and
widths are simply determined as the position and full width at half-maximum of the
resonant peak, respectively. With this method, one can describe narrow resonances very
well, but the accuracy is poor for wide ones. Therefore, in our recent
studies~\cite{PRC2020Sun_101_014321,CPC2020Chen44_084105}, we proposed an effective and direct way
to identify the resonant states by exploring for the extremum of the DOS. The exact energies and widths for the resonant states in all types can be obtained, whether
for wide or narrow resonant states. However, the DOS in the
calculations are approximate ones because they are calculated in a finite space size. In
this work, we will directly analyze Green's functions and search for their poles or
extrema to determine the resonant states.

The paper is organized as follows: The RMF model formulated with Green's functions is
briefly presented in Sec.~\ref{sec:Theory}. Numerical details are given in
Sec.~\ref{sec:Number}. After the results and discussion are presented in Sec.~\ref{sec:Resu},
a brief summary and perspectives are given in Sec.~\ref{sec:Sum}.

\section{THEORETICAL FRAMEWORK}
\label{sec:Theory} In the RMF model, neutrons and protons are described as Dirac
particles moving in a mean-field potential characterized by scalar $S$ and vector
$V$ potentials. The Dirac equation for a nucleon with mass $M$ is as follows:
\begin{equation}
[\bm{\alpha}\cdot\bm{p}+V(\bm{r})+\beta(M+S(\bm{r}))]\psi_n(\bm{r})=\varepsilon_n\psi_n(\bm{r}),
\end{equation}
where $\bm{\alpha}$ and $\beta$ are Dirac matrices.

Various methods have been used to solve the Dirac equation. These include the shooting
method~\cite{NPA1998JMeng_635_3}, the Green's function method~\cite{PRC2014TTSun_90_054321}, and the
finite element method~\cite{NST2020JYFang}, which are performed in the coordinate space,
as well as those in the harmonic oscillator basis~\cite{NY1990YKGambhir} or Woods--Saxon
basis~\cite{PRC2003SGzhou}. When introducing the Green's function
method~\cite{Book2006Eleftherios-GF} to mean-field density functionals, the densities
and single-particle spectrum can be determined directly by the Green's
functions~\cite{PRC2011ZhangY_83_054301,PRC2012YZhang_86_054318,PRC2014TTSun_90_054321,PRC2019SunTT_99_054316}.
Following the definition of the single-particle Green's function,
\begin{equation}
[\varepsilon-\hat{h}(\bm{r})]\mathcal{G}(\bm{r},\bm{r}';\varepsilon)=\delta(\bm{r}-\bm{r}'),
\label{Eq:GFdef}
\end{equation}
a relativistic Green's function $\mathcal{G}(\bm{r},\bm{r'};\varepsilon)$ for the Dirac
equation can be constructed at arbitrary single-particle energies $\varepsilon$ when
$\hat{h}(\bm{r})$ is chosen as the Dirac Hamiltonian. Taking a complete set of solutions of
the Dirac equation, including the eigenstates $\psi_{n}(\bm{r})$ and eigenvalues
$\varepsilon_{n}$, we can write the Green's function in Eq.~(\ref{Eq:GFdef})
as
\begin{equation}
\mathcal{G}(\bm{r},\bm{r}';\varepsilon)=\sum_n\frac{\psi_{n}(\bm{r})\psi_{n}^{\dag}(\bm{r}')}{\varepsilon-\varepsilon_{n}}.
\label{Eq:GFdef2}%
\end{equation}%
With the single-particle energy $\varepsilon$ approaching the energy $\varepsilon_n$,
the absolute value of $\mathcal{G}(\bm{r},\bm{r}';\varepsilon)$ will increase
significantly and reach the extremum. Therefore, one can determine the single-particle
energies $\varepsilon_n$ by calculating different Green's functions at various energies
$\varepsilon$ and search for the extremum. For resonant states with  resonant
energies $E$ and widths $\Gamma$, one can write their energies as
$\varepsilon_{n}=E-i\Gamma/2$. Correspondingly, the energies $\varepsilon$ in
Eqs.~(\ref{Eq:GFdef}) and (\ref{Eq:GFdef2}) are complex:
$\varepsilon=\varepsilon_r+i\varepsilon_i$ with $\varepsilon_r$ and $\varepsilon_i$
being the real and imaginary parts of the energy, respectively.

Because the Dirac spinor $\psi_{n}(\bm{r})$ has upper and lower components, the corresponding Green's function has the form of a $2\times2$ matrix,
\begin{equation}
\mathcal{G}(\bm{r},\bm{r}';\varepsilon)=
\left(
 \begin{array}{cc}
  \mathcal{G}^{(11)}(\bm{r},\bm{r}';\varepsilon) & \mathcal{G}^{(12)}(\bm{r},\bm{r}';\varepsilon) \\
  \mathcal{G}^{(21)}(\bm{r},\bm{r}';\varepsilon) & \mathcal{G}^{(22)}(\bm{r},\bm{r}';\varepsilon)
 \end{array}
\right).
\label{Eq:GFm}
\end{equation}

With spherical symmetry, one can expand the Green's function as
\begin{equation}
\mathcal{G}({\bm r},{\bm r'};\varepsilon)=\sum_{\kappa m}Y_{\kappa m}(\theta,\phi)\frac{\mathcal{G}_{\kappa}(r,r';\varepsilon)}{rr'}Y_{\kappa m}^{*}(\theta',\phi'),
\end{equation}
where $\mathcal{G}_{\kappa}(r,r';\varepsilon)$ is the radial part, $Y_{\kappa
m}(\theta,\phi)$ is the spin spherical harmonic, and the quantum number
$\kappa=(-1)^{j+l+1/2}(j+1/2)$ labels different ``channels.''

For a given single-particle energy $\varepsilon$ and quantum number $\kappa$, we can
construct the radial Green's function $\mathcal{G}_{\kappa}(r, r';\varepsilon)$
as~\cite{PRB1992Tamura_45_3271}
\begin{flalign}
\mathcal{G}_{\kappa}(r,r';\varepsilon)=
&\frac{1}{W_{\kappa}(\varepsilon)}\left[\theta(r-r')\phi^{(2)}_{\kappa}(r,\varepsilon)\phi^{(1)\dag}_{\kappa}(r',\varepsilon)\right. \nonumber\\
&\left.+\theta(r'-r)\phi^{(1)}_{\kappa}(r,\varepsilon)\phi^{(2)\dag}_{\kappa}(r',\varepsilon)\right],
\label{Eq:GFcons}
\end{flalign}
with $\phi^{(1)}_{\kappa}(r,\varepsilon)$ and $\phi^{(2)}_{\kappa}(r,\varepsilon)$ being two Dirac spinors given by
\begin{eqnarray}
&&\phi^{(1)}_{\kappa}(r,\varepsilon)=\left( \begin{array}{c}
g^{(1)}_{\kappa}(r,\varepsilon) \\
f^{(1)}_{\kappa}(r,\varepsilon)
\end{array} \right),\nonumber\\
&&\phi^{(2)}_{\kappa}(r,\varepsilon)=\left( \begin{array}{c}
g^{(2)}_{\kappa}(r,\varepsilon) \\
f^{(2)}_{\kappa}(r,\varepsilon)
\end{array} \right),
\end{eqnarray}
which are linearly independent and obtained by solving the Runge--Kutta integrals in the
full coordinate $r$ space starting, respectively, from the asymptotic behaviors at
$r\rightarrow0$ and $r\rightarrow\infty$; $\theta(r-r')$ is a step function; and
$W_{\kappa}(\varepsilon)$ is the Wronskian function,
\begin{equation}
W_{\kappa}(\varepsilon)=g^{(1)}_{\kappa}(r,\varepsilon)f^{(2)}_{\kappa}(r,\varepsilon)-g^{(2)}_{\kappa}(r,\varepsilon)f^{(1)}_{\kappa}(r,\varepsilon),
\end{equation}
which is independent of $r$. It can be checked that the constructed Green's function
$\mathcal{G}_{\kappa}(r,r';\varepsilon)$ of Eq.~(\ref{Eq:GFcons}) meets the definition
equation~(\ref{Eq:GFdef}) in the radial form.

\begin{figure*}[tp!]
\includegraphics[width=0.65\textwidth]{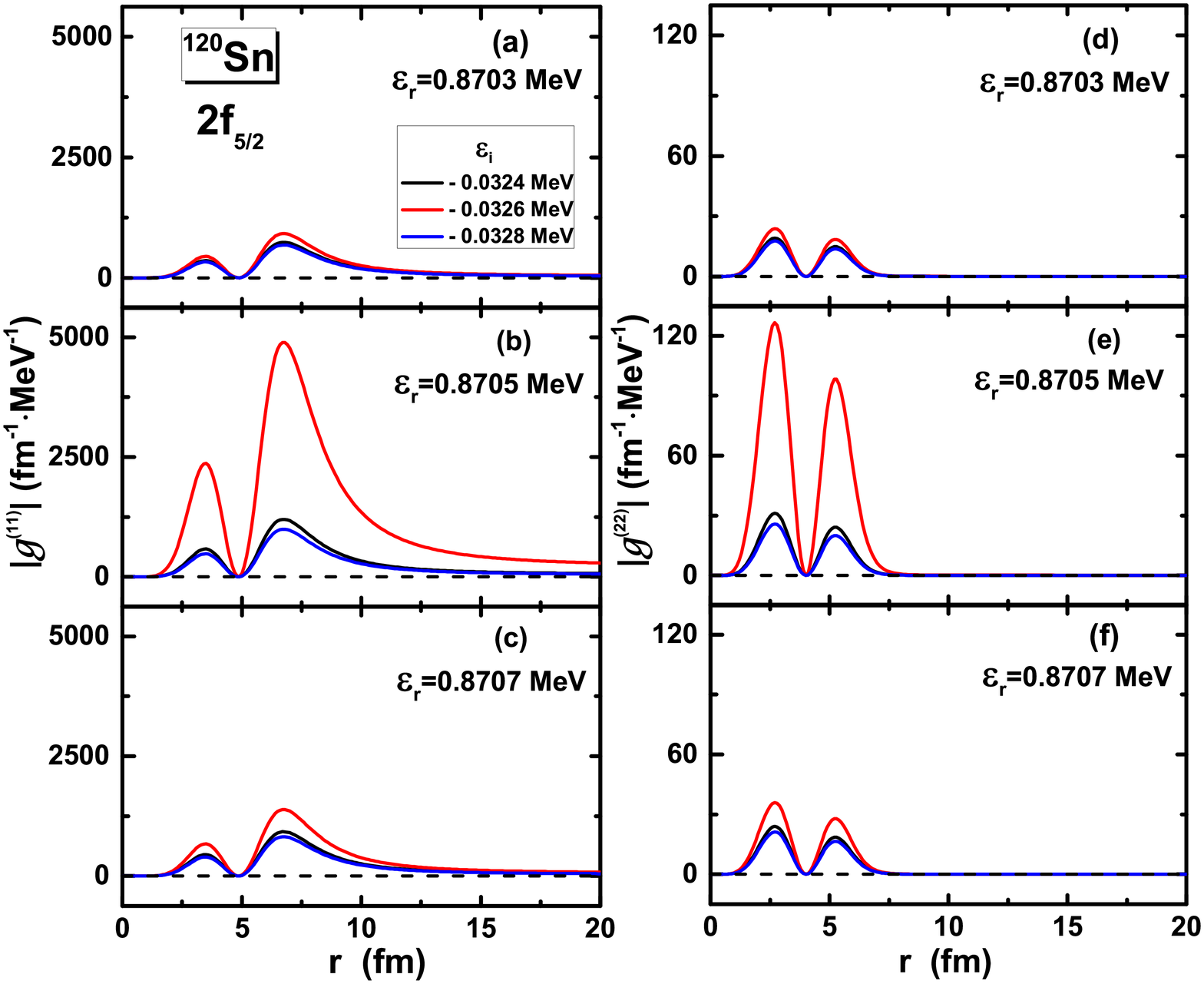}
\caption{ (Color online) Green's functions $\mathcal{G}(r,r;\varepsilon)$ at different complex energies $\varepsilon=\varepsilon_r+i\varepsilon_i$ plotted as a function of coordinate $r$ for the resonant state $2f_{5/2}$ in $^{120}$Sn. The left and right columns are, respectively, the moduli of ``11" and ``22" components of Green's functions $|\mathcal{G}^{(11)}|$ and $|\mathcal{G}^{(22)}|$. The complex energy $\varepsilon$ is scanned widely, and the results with
the real energies $\varepsilon_r=0.8703,$ $0.8705$, and $0.8707$~MeV, and imaginary energies $\varepsilon_i=-0.0324$, $-0.0326$, and $-0.0328$~MeV are shown.
}
\label{Fig1}
\end{figure*}

In practical calculations, one will adopt the exact asymptotic behaviors of the
Dirac spinors to construct Green's functions. As a result, weakly bound states around the Fermi surface as well as the resonances above the continuum threshold, which
are essential for the unstable nuclei, can be adddressed when calculating
densities and single-particle spectra. At $r\rightarrow 0$, the asymptotic behavior of the
Dirac spinor $\phi_{\kappa}^{(1)}(r,\varepsilon)$ satisfies
\begin{eqnarray}
\phi^{(1)}_{\kappa}(r,\varepsilon)
   &\longrightarrow& r\left(
                                                       \begin{array}{c}
                                                         j_l(k r) \\
                                                         \frac{\kappa}{|\kappa|}\frac{\varepsilon-V-S}{k}j_{\tilde{l}}(kr)\\
                                                       \end{array}
                                                     \right),
                                                  \label{Eq:behavior_r0}
 \end{eqnarray}
where $k=\sqrt{(\varepsilon-V-S)(\varepsilon-V+S+2M)}$ is the single-particle momentum,
the quantum number $\tilde{l}=l+(-1)^{j+l-1/2}$ denotes the angular momentum for the
lower component of the Dirac spinor, and $j_l(k r)$ is the spherical Bessel
function of the first kind, which satisfies
\begin{equation}
j_l(k r)\longrightarrow \frac{(kr)^l}{(2l+1)!!},~~~\text{ when}~~r\rightarrow 0.
\end{equation}

At $r\rightarrow\infty$, the Dirac spinor $\phi_{\kappa}^{(2)}(r,\varepsilon)$ is
oscillating outgoing for the continuum and exponentially decaying for the bound states,
which can be represented uniformly as
\begin{eqnarray}
\phi^{(2)}_{\kappa}(r,\varepsilon)
 &\longrightarrow&\left(
                                                    \begin{array}{c}
                                                      rk h^{(1)}_l(k r) \\
                                                      \frac{\kappa}{|\kappa|}\frac{rk^2}{\varepsilon+2M}h^{(1)}_{\tilde{l}}(k r) \\
                                                    \end{array}
                                                  \right),
\label{Eq:behavior_rinf}
\end{eqnarray}
where $k=\sqrt{\varepsilon(\varepsilon+2M)}$ and $h^{(1)}_l(k r)$ is the spherical Hankel function of the first kind.

\section{Numerical details}
\label{sec:Number}

In this work, the single-particle resonant states were studied by employing the Green's
function method based on RMF theory, where the resonance energies $E$ and widths
$\Gamma$ were obtained directly by probing the extrema of the Green's functions. To compare
with the previous results obtained by using Green's functions
calculations~\cite{PRC2014TTSun_90_054321,CPC2020Chen44_084105}, where the resonances
were identified by probing the extremum of the DOS,
$n_{\kappa}(\varepsilon)$, defined in a finite space size $R_{\rm box}$, the same nucleus
$^{120}$Sn and density functional PK1~\cite{PRC2010Zhao_82_054319} were adopted.

The Green's functions and RMF equations were solved in the coordinate $r$ space with a
space size of $R_{\rm{box}}=20$~fm and a step of $dr=0.1$~fm. To check the dependence of
the obtained resonances on the space sizes, calculations with $R_{\rm{box}}=25, 30$~fm
were also performed. To search for the energy position corresponding to the poles of the
Green's functions, the complex single-particle energies
$\varepsilon=\varepsilon_r+i\varepsilon_i$ were scanned widely on the complex energy
plane to calculate the Green's functions $\mathcal{G}(r,r;\varepsilon)$. The energy step was
$d\varepsilon=1\times 10^{-4}$~MeV for both the real~$\varepsilon_r$~and
imaginary~$\varepsilon_i$~parts. With this scanning energy step, the accuracy of the
obtained resonance energies and widths were up to $0.1$ keV. In addition, the accuracy can
be increased further once the scanning energy step $d\varepsilon$ is decreased.

\section{Results and discussion}
\label{sec:Resu}

On the complex single-particle energy plane, single-particle resonances are located in
the fourth quadrant, with the real energy being the resonance energy $E$ and the imaginary
energy being half of the resonance width $\Gamma/2$. According to Eq.~(\ref{Eq:GFdef2}),
the single-particle energies $\varepsilon_n=E-i\Gamma/2$ of the resonant states also
correspond to the extrema of the Green's functions $\mathcal{G}(r,r;\varepsilon)$. As a
result, one can search for these poles or extrema to determine the locations of the
resonate sates by scanning the complex energies
$\varepsilon=\varepsilon_r+i\varepsilon_i$ in the fourth quadrant and calculating the Green's
functions $\mathcal{G}(r,r;\varepsilon)$.

\begin{figure}[tp!]
\includegraphics[width=0.45\textwidth]{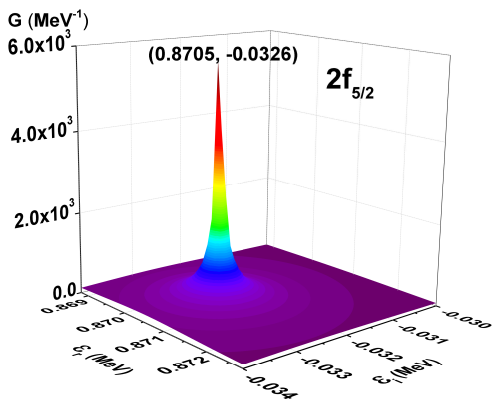}
\caption{(Color online) Integral function $G_{\kappa}(\varepsilon)$ distributed on the complex energy plane for the resonant state $2f_{5/2}$ in $^{120}$Sn.}
\label{Fig2}
\end{figure}

In Fig.~\ref{Fig1}, taking the neutron single-particle resonant state $2f_{5/2}$ in
$^{120}$Sn as an example, we plot the Green's functions $\mathcal{G}(r,r;\varepsilon)$ at
various scanned single-particle complex energies
$\varepsilon=\varepsilon_r+i\varepsilon_i$. With the real energy
$\varepsilon_{r}$ varying from $0.8703$ to $0.8707$~MeV and the imaginary energy
$\varepsilon_i$ varying from $-0.0324$ to $-0.0328$~MeV, the height of the Green's
function changes significantly. Comparing panels (a), (b), and (c) in the left
column, we can see that the modulus of the Green's function
$|\mathcal{G}^{(11)}|$ has a larger amplitude at a real energy
$\varepsilon_r=0.8705~$MeV, as shown in panel (b). More specifically, the Green's function
reaches its extremum at an imaginary energy $\varepsilon_i=-0.0326~$MeV (plotted by the red
line). All these analyses indicate that the Green's function reaches its extremum at an
energy of $\varepsilon=0.8705-i0.0326$~MeV. In the same way, we show the moduli of the
Green's functions for the ``22" component $|\mathcal{G}^{(22)}|$ in the right column,
which are determined by the small component of the Dirac spinor. The amplitudes are much
lower than those in the left column. However, the Green's function reaches its
maximum amplitude at the same energy $\varepsilon=0.8705-i0.0326$~MeV. Therefore, we can
conclude that $\varepsilon=0.8705-i0.0326$~MeV corresponds to the energy of the
single-particle resonant state $2f_{5/2}$ in $^{120}$Sn, which is almost the same
value as the result obtained by exploring for the extremum of the DOS ~\cite{CPC2020Chen44_084105} with a difference of $0.1~$keV for the width.

To be more intuitive, one can integrate the Green's function
$\mathcal{G}(r,r;\varepsilon)$ over coordinate $r$ and compare the integral values
at different scanning single-particle energies $\varepsilon$. The integral function
$G_{\kappa}(\varepsilon)$ for each partial wave $\kappa$ at energy $\varepsilon$ can be
written as
\begin{equation}
{G}_{\kappa}(\varepsilon)=\int dr[|\mathcal{G}_{\kappa}^{(11)}(r,r;\varepsilon)|+|\mathcal{G}_{\kappa}^{(22)}(r,r;\varepsilon)|],
\label{Eq:GFiter}
\end{equation}
where $|\mathcal{G}_{\kappa}^{(11)}(r,r;\varepsilon)|$ and
$|\mathcal{G}_{\kappa}^{(22)}(r,r;\varepsilon)|$, respectively, correspond to the moduli
of the ``11'' and ``22'' matrix elements of the Green's functions, as shown in
Eq.~(\ref{Eq:GFm}). A sharp peak should be observed for the integral function
$G_{\kappa}(\varepsilon)$ at the energy where the Green's function reaches its extremum.
In Fig.~\ref{Fig2}, the result for the single-particle resonant state $2f_{5/2}$ in
$^{120}$Sn is plotted on the complex energy plane. A peak with an extremum located at
$\varepsilon_r=0.8705$~MeV and $\varepsilon_i=-0.0326$~MeV is observed, indicating that the
energy of the resonant state $2f_{5/2}$ is $\varepsilon_n=0.8705-i0.0326$~MeV, which is
the same as the result obtained in Fig.~\ref{Fig1}.

\begin{figure}[tp!]
\includegraphics[width=0.45\textwidth]{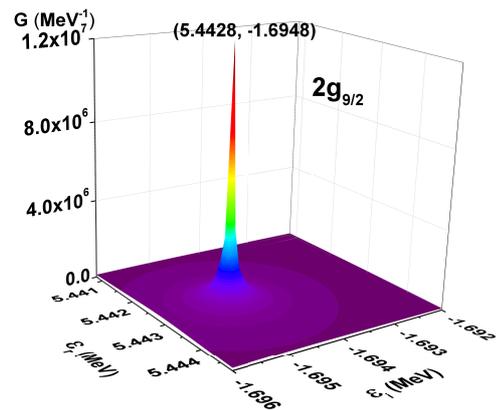}
\caption{(Color online) Integral function $G_{\kappa}(\varepsilon)$ distributed on the complex energy plane for the resonant state $2g_{9/2}$ in $^{120}$Sn.}
\label{Fig3}
\end{figure}

\begin{figure}[htp!]
\includegraphics[width=0.35\textwidth]{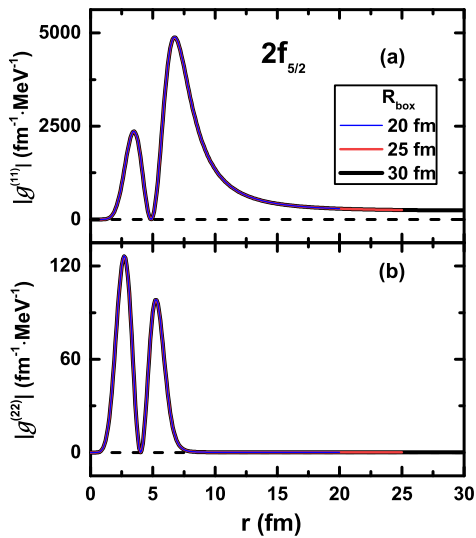}
\caption{ (Color online) Green's functions $\mathcal{G}(r,r;\varepsilon)$ for the resonant state $2f_{5/2}$ in $^{120}$Sn at the resonant energy $\varepsilon=E-i\Gamma/2$ calculated with space sizes $R_{\rm box}=$20, 25, and~30~fm.}
\label{Fig4}
\end{figure}

\begin{table*}[tp!]
\caption {Neutron single-particle resonances $nl_j$ in $^{120}$Sn with energies and widths $E-i\Gamma/2$ obtained by probing the extrema of the Green's functions, compared with the results by exploring for the extremum of the DOS in our previous study~\cite{CPC2020Chen44_084105}. The PK1 effective density functional was used. All quantities are in MeV. }
\begin{center}
{
\begin{tabular}{ccc||ccc}
\hline
Positive parity  &             Green's function      & DOS               & Negative parity &                   Green's function      & DOS      \\
\hline
$2g_{7/2}$       & $6.3585- i3.1053$   & $6.3585- i3.1052$ & $3p_{1/2}$      & $0.0504- i0.0164$          & $0.0504- i0.0164$ \\
$2g_{9/2}$       & $5.4428- i1.6948$   & $5.4428- i1.6948$ & $2f_{5/2}$      & $0.8705- i0.0326$          & $0.8705- i0.0325$ \\
$1i_{11/2}$      & $9.8544-i0.6413$    & $9.8544- i0.6413$ & $1h_{9/2}$      & $0.2507- i4\times10^{-8}$  & $0.2508- i4\times10^{-8}$ \\
$1i_{13/2}$      & $3.4686-i0.0025$    & $3.4686-i0.0024$  & $2h_{11/2}$     & $10.5130-i6.7683$         & $10.5130- i6.7681$ \\
                                                        && & $1j_{13/2}$     & $18.1846-i3.1532$         & $18.1846- i3.1531$ \\
                                                        && & $1j_{15/2}$     & $12.8929-i0.5323$         & $12.8929- i0.5322$ \\
\hline
\end{tabular}}
\end{center}
\label{Tab1}
\end{table*}

To check the universality of this approach, the same analysis was performed for a
relatively wide single-particle resonant state $2g_{9/2}$ in Fig.~\ref{Fig3}. Similarly,
a sharp peak is identified with the extremum located at
$\varepsilon_r=5.4428$~MeV and $\varepsilon_i=-1.6948$~MeV, indicating that the energy of the
resonant state $2g_{9/2}$ is $\varepsilon_n=5.4428-i1.6948$~MeV. However, compared with
the narrow resonant state $2f_{5/2}$, the peak of the wide resonant state $ 2 g _{9/2}$
is much sharper, which can be explained by the greater integral values
$G_{\kappa}(\varepsilon)$ for the wide resonant states caused by the extended
distributions of Green's functions. From Figs.~\ref{Fig2} and \ref{Fig3}, we can
conclude that it is very direct and effective to search for the single-particle
resonant states and determine the energies and widths for both narrow and wide
resonances by searching for the extrema of the Green's functions.

In Fig.~\ref{Fig4}, the dependence of the Green's functions on the coordinate space
sizes $R_{\rm box}$ is checked by taking the resonant state $2f_{5/2}$ as an example and
plotting  $|\mathcal{G}^{(11)}(r,r;\varepsilon)|$ and
$|\mathcal{G}^{(22)}(r,r;\varepsilon)|$ with $R_{\rm box}=20 $, $ 25$, and $30$~fm, respectively.
Obviously, we can see that exactly the same distributions for both the ``11" and ``22"
components are obtained, indicating that the Green's function method
depends slightly on the space sizes, which is consistent with the conclusions obtained by
analyzing the density distributions $\rho_{\kappa}(r,\varepsilon)$ plotted in different
space sizes~\cite{CPC2020Chen44_084105}.

In Table~\ref{Tab1}, neutron single-particle resonant states in $^{120}$Sn  with
energies and widths $E-i\Gamma/2$ obtained by searching for the extrema of the Green's
functions are listed, and these are compared with those obtained by exploring for the maximum of the
DOS~\cite{CPC2020Chen44_084105}. For most of the resonant sates, exactly
the same energies and widths are obtained while very small differences (within $0.2$~keV)
exist for others. Compared with the results obtained by comparing the DOSs for nucleons and free particles in Ref.~\cite{PRC2014TTSun_90_054321}, four wide
resonant states ($ 2 g _{7/2}$, $ 2 g _{9/2}$, $2h_{11/2}$, and $1j_{13/2}$) are identified.
Furthermore, the accuracy of the width for the narrow resonant state $1h_{9/2}$ is
highly refined to be $1\times 10^{-8}$~MeV. All these results prove that the approach by
probing the extremum of the Green's functions is as effective and reliable as that by exploring for the extremum of the DOS in identifying the resonant
states, irrespective of whether the resonant state is wide or narrow. Moreover, compared with the
approach by exploring for the extremum of the DOS, the approach by searching
for the extremum of the Green's functions is easier, more direct, and less time-consuming.
However, without the DOS, this approach cannot describe intuitively the
structures of the single-particle spectrum for the bound and resonant states.

\section{Summary and Perspectives}
\label{sec:Sum} Significant roles are played by the single-particle resonances in the
structure of exotic nuclei. The Green's function method has been demonstrated to be one of
the most effective approaches in searching for single-particle resonant states. In our
recent work~\cite{CPC2020Chen44_084105}, by probing the extremum of the DOSs
$n_{\kappa}(\varepsilon)$ defined in a finite space size $R_{\rm box}$, the Green's function
method has been proven to be very reliable, regardless of whether the resonances are wide or narrow. In
this work, another direct and effective approach by probing the extremum of the Green's
functions is proposed to identify the resonant states.

Taking the same nucleus $^{120}$Sn as an example, by searching for the poles or extrema
of the Green's functions, we obtain almost the same energies and widths for the resonant
states as obtained by exploring for the extremum of the DOS. In addition, the
dependence of the Green's functions on the space size is checked and found to be very
stable. Compared with the results obtained by comparing the DOSs for
nucleons and free particles~\cite{PRC2014TTSun_90_054321}, four wide resonant states
($ 2 g _{7/2}$, $ 2 g _{9/2}$, $2h_{11/2}$, and $1j_{13/2}$) are identified, and the accuracy of
the width of the narrow resonant state $1h_{9/2}$ is highly refined to $1\times
10^{-8}$~MeV. All these results prove that the approach by probing the extremum of the Green's
functions has the same reliability and effectiveness as that by probing the
extremum of the DOS to identify the resonant states, regardless of whether the resonant states are wide or
narrow.

As is well known, both pairing correlations and the continuum play core roles in exotic
nuclei. Therefore, studies on the possible effects of pairing on the resonant states are
significant and very interesting. In the investigations with the continuum Skyrme HFB
approach in Refs.~\cite{SC2019QuXY,PRC2020ZhangY}, the authors concluded that the pairing correlation can
enhance the resonant energies and widths for all quasiparticle resonances, whether
hole-like or particle-like. However, in
Ref.~\cite{PTEP2016Kobayashi}, an opposite conclusion was obtained for the particle-like
quasiparticle resonances studied with a fixed resonant energy. In the future, we will
take the self-consistent relativistic continuum Hartree--Bogoliubov model with the Green's
function method~\cite{Sci2016Sun_46_12006} to explore the possible effects of pairing on
the resonant states. Moreover, we would like to apply the Green's function method to dynamic reactions and search for resonance structures, for which vast
theoretical and experimental works have been performed~\cite{NST2019HRGuo,NST2019WJLi}.
In Ref.~\cite{NST2019XDTang}, the complex molecular resonances in the $^{12}{\rm C}+^{12}$C
fusion reaction were explored with the thick-target method.

\begin{acknowledgements}
Helpful discussions with Prof. Z.~P. Li are highly appreciated.
\end{acknowledgements}

\end{document}